\documentclass[aps,twocolumn,pra,showpacs,amsmath,amssymb]{revtex4}

\usepackage{graphicx}

\def\be{\begin{equation}}
\def\ee{\end{equation}}
\def\bea{\begin{eqnarray}}
\def\eea{\end{eqnarray}}
\def\bma{\begin{mathletters}}
\def\ema{\end{mathletters}}

\def\0{\overline{0}}

\def\q0{\underline{0}}

\def\tr{\mbox{tr}}
\def\one{\leavevmode\hbox{\small1\normalsize\kern-. 33em1}}

\begin{document}

\title{Secret key distillation from shielded two-qubit states}

\author{Joonwoo Bae}
\affiliation{School of Computational Sciences, Korea Institute for
Advanced Study, Seoul 130-722, Korea}

\date{\today}


\begin{abstract}
The quantum states corresponding to a secret key are characterized by the so-called private states, where the key part consisting of a secret key is shielded by the additional systems. Based on the construction, it was shown that a secret key can be distilled from bound entangled states. In this work, we consider the shielded two-qubit states in key distillation scenario and derive the conditions that a secret key can be distilled using the recurrence protocol or the two-way classical distillation, the advantage distillation together with one-way post-processing. From the security conditions, it is shown that a secret key can be distilled from bound entangled states in a much wider range. In addition, we consider the case that the white noise is added to quantum states, and show that the classical distillation protocol still works under certain amount of the noise although the recurrence protocol not.
\end{abstract}


\maketitle

\section{Introduction}

Entanglement, i.e. quantum correlations that cannot be prepared by local operations and classical communication (LOCC), has been the resource for efficient quantum information processing. One of the important applications is that entangled states can be used to distribute a secret key \cite{skey}. For instance, in the Bennett-Brassard (BB84) protocol \cite{bb84}, if entangled states shared by two honest parties are not too noisy, then they can be used to establish a secret key between two honest parties, Alice and Bob, applying single-copy level measurement and one-way post-processing for error correction and privacy amplification. It is the key idea in the security proof that distilling the maximally entangled state, $|\phi_{1}\rangle = (|00\rangle + |11\rangle)/ \sqrt{2}$, immediately implies that a secret key is obtained \cite{sp}. This is based on two characteristics of the entangled state. First, since the state is pure, two honest parties do not share any correlation with environment, i.e. the probability distribution with an eavesdropper is of the form, $P_{ABE} (a,b,e) = P_{AB}(a,b) P_{E}(e)$. Second, measuring copies of the state in the computational basis, two parties share the perfect correlation, i.e. $P_{AB}(0,0) = P_{AB}(1,1) =1/2$.

If the limitation on technology of Alice and Bob is relaxed, i.e. two honest parties are assumed to apply coherent quantum operations over shared quantum states, then the so-called distillable entangled states \cite{entanglement distillation}, those entangled states that can be distilled to the maximally entangled states by local operation and classical communication(LOCC), can be used to establish a secret key between honest parties. Then, one can ask if the entanglement distillability is the characterization of key-distillability? Or, more generally, for given quantum states, is it possible to decide if a secret key can be distilled? This has been referred to as one of the open problems in Quantum Information Theory, being a fundamental issue on the compatibility between entanglement and secrecy \cite{prob 24}. Along this line, it is clear that separable states cannot be used in secret key distillation \cite{CLL,sc}. It is also known that all entangled states can be used to establish classical secret correlations, those classical correlations that cannot be prepared by local operation and public communication \cite{CLL, sc}.

In Ref. \cite{horodecki}, the quantum states that correspond to a secret key are characterized by private states $\gamma_{ABA'B'}$ as follows, \bea \gamma_{ABA'B'} = U_{ABA'B'} (|\phi_{1}\rangle_{AB} \langle \phi_{1}|\otimes \rho_{A'B'})U_{ABA'B'}^{\dagger}, \label{pbit} \eea where $\rho_{A'B'}$ can be any quantum state, and $U_{ABA'B'}$ is a unitary operation called \emph{twisting} of the following form \bea U_{ABA'B'} = \sum_{i,j} |ij\rangle_{AB} \langle ij|\otimes U_{ij}^{A'B'}. \label{twisting}\eea A secret key can be obtained by measuring the systems $A$ and $B$, which are called the key part that are directly used to distilling a secret key. The other systems $A'$ and $B'$ are called shield parts that are only used to shield the key part. It can be now said that the key-distillable states are those quantum states that can be distilled to private states by LOCC. Based on the private states distillation scenario, it was shown that there exist bound entangled states from which a secret key can be distilled \cite{horodecki}. This was the first line that discovers the incompatibility between entanglement and secrecy properties of quantum states. Therefore, it turns out that the set of key-distillable states is strictly larger than that of distillable entangled states.

The bound entangled states constructed in \cite{horodecki} is of the following form, \bea \rho_{ABA'B'}&=& |\phi_{1}\rangle\langle\phi_{1}|_{AB}\otimes\sigma_{1} +
|\phi_{2}\rangle\langle\phi_{2}|_{AB}\otimes\sigma_{2} \nonumber\\
&+&|\phi_{3}\rangle\langle\phi_{3}|_{AB}\otimes\sigma_{3}
+|\phi_{4}\rangle\langle\phi_{4}|_{AB}\otimes\sigma_{4},
\label{state} \eea where $\sigma_{j}$, $j=1,2,3,4$, are unnormalized shield states of systems $A'B'$ and $|\phi_{i}\rangle$, $j=1,2,3,4$ are Bell states, $|\phi_{2}\rangle = (|00\rangle - |11\rangle)/\sqrt{2}$, $|\phi_{3}\rangle = (|01\rangle + |10\rangle)/\sqrt{2}$ and $|\phi_{4}\rangle = (|01\rangle + |10\rangle)/\sqrt{2}$. Throughout, we call the state in (\ref{state}) as shielded two-qubit states. The shield states in Ref.\cite{horodecki} are separable and constructed such that the state $\rho_{ABA'B'}$ is of non-positive-partial-transpose (NPPT). The recurrence protocol in Ref. \cite{entanglement distillation} that was used to distill maximally entangled states was applied to the key parts of the state $\rho_{ABA'B'}$ in order to distill a secret key \cite{horodecki}. Since the state is of NPPT it is clear that no maximally entangled state can be distilled, however, it was shown that the resulting state converges to a private state by the recurrence protocol.

In this paper, we consider the shield two-qubit states in key distillation scenario. We apply two distillation techniques, the recurrence protocol, and the two-way classical communication protocol, the advantage distillation together with the standard one-way post-processing for error-correction and privacy amplification. We then derive the conditions that a secret key can be distilled by the quantum and classical protocols, respectively, and show that bound entangled sates in a much wider range are turned out to be key-distillable. We also show that the classical distillation protocol is robust when the white noise is added, while the recurrence protocol not.

\section{Two-qubits with shield systems}
\label{shiled}

Let us begin with reviewing properties of private states in (\ref{state}), where two honest parties hold both systems: one is the key part $AB$ to be directly used for key distillation, and the other is the shield part $A'B'$. In fact, all private states are equivalent up to twisting operations. In Ref. \cite{other}, it was shown that for $\rho_{ABA'B'}$ in (\ref{state}), there exists a twisting $U$ such that $\rho_{ABA'B'} =
U\sigma_{ABA'B'} U^{\dagger}$ in which the key
part, $\sigma_{AB} = \tr_{A'B'}[U\sigma_{ABA'B'} U^{\dagger}]$, becomes \bea \sigma_{AB} & = &
\lambda_{1}|\phi_{1}\rangle\langle\phi_{1}| +
\lambda_{2}|\phi_{2}\rangle\langle\phi_{2}| \nonumber \\ &+&
\lambda_{3}|\phi_{3}\rangle\langle\phi_{3}| +
\lambda_{4}|\phi_{4}\rangle\langle\phi_{4}|, \label{p-state} \eea
with \bea \lambda_{1,2} & = & \frac{1}{2}(\|\sigma_{1} +\sigma_{2}
\| \pm \|\sigma_{1}
-\sigma_{2} \|),\nonumber \\
\lambda_{3,4} & = & \frac{1}{2}(\|\sigma_{3} +\sigma_{4} \| \pm
\|\sigma_{3}-\sigma_{4} \|).\label{lam} \eea Note that there exists an LOCC scheme to estimates the parameters $\lambda_{i}$($i=1,2,3,4$) in the above \cite{pri}.

The twisting operations could be in general non-local quantum operations. They are, however, useful to analyze the so-called classical-classical-quantum(ccq) correlations of Alice, Bob and Eve when Alice and Bob measure their key parts in the computational basis $\{ |0\rangle,|1\rangle \}$ and obtain classical raw keys. For the full generality, Eve is supposed to be able to postpone measurement. The important property of the twisting operations is that, the ccq correlations among them do not depend on application of twisting operations, i.e. the ccq correlations are the same for both states, $\rho_{ABA'B'}$ and the twisted one $U\rho_{ABA'B'}U^{\dagger}$  \cite{other, pri}. \\

\textbf{Theorem} (\cite{other}). For any state $\rho_{ABA'B'}$ of the form in (\ref{state}) and any twisting operation $U$, the states $\rho_{ABA'B'}$ and $U\rho_{ABA'B'} U^{\dagger}$ have the same ccq correlations with respect to measurement in the computational basis. That is, the corresponding ccq states are equivalent. \\

Therefore, when sharing copies of the state $\rho_{ABA'B'}$, if two honest parties measure the key parts and apply classical distillation protocol, the security can be equivalently analyzed with the twisted one $U \rho_{ABA'B'} U^{\dagger}$, in order to decide whether the state $\rho_{ABA'B'}$ is key-distillable. The precondition for key distillability \cite{CLL,sc} is that the key part $\sigma_{AB}$ is entangled for some $U$. For the twisting $U$ achieving (\ref{p-state}) and (\ref{lam}), the key part is entangled if and only if $\lambda_{1} > \lambda_{2} + \lambda_{3} + \lambda_{4}$ \cite{ent}. This can be equivalently expressed in terms of shield states (\ref{lam}) \bea \|\sigma_{1}-\sigma_{2} \| > \|\sigma_{3} + \sigma_{4} \|. \label{ent}\eea

In the key distillation scenario, we will consider the case of so-called collective attacks, in which the shared states of two honest parties are $N$ copies of the same states, $\rho_{ABA'B'}^{\otimes N}$. That is, the collective attacks correspond to the case that Eve interacts with individual quantum state of two honest parties. This, however, does not lose the generality in proving the security against general attacks, i.e. the general security, in the asymptotic limit $N\rightarrow \infty$. Thanks to the quantum de Finetti theorem, it turns out that the general attacks are not more powerful than the collective attacks for sufficiently large $N$ \cite{thesis}. The security analysis can therefore be hugely simplified to the case of collective attacks.

\section{Distilling key by the recurrence protocol}

We now apply the recurrence protocol to distill a secret key from  $\rho_{ABA'B'}^{\otimes N}$. The recurrence protocol applies only to the key parts and works as follows \cite{entanglement distillation}. Two honest parties first take two copies of the state, $\rho_{A_{1}B_{1}A'_{1}B'_{1}} \otimes \rho_{A_{2}B_{2}A'_{2}B'_{2}}$. The CNOT operations are applied to the local systems $A_{1}A_{2}$ and $B_{1}B_{2}$, respectively, where $A_{1}$ and $B_{1}$ are controlled systems and $A_{2}$ and $B_{2}$ are target ones. Then, two honest parties measure the key parts of the second pair, $A_{2}$ and $B_{2}$, in the computational basis, and communicate if their measurement outcomes are the same. If measurement outcomes are the same, two honest parties keep the resulting quantum state and repeat the procedure with the remaining pairs. If not, they start the protocol with another pairs. Let $\sigma_{m}$ denote the resulting state after passing $m$ repetitions of the protocol. If there exists a private state to which $\sigma_{m}$ converges as $m$ tends to be large, it is concluded that $\rho_{ABA'B'}$ is key-distillable. In Ref. \cite{horodecki,other}, the useful relation was shown that, the state $\sigma_{m}$ converges to a private state as $m$ becomes large if and only if  $\| \langle 00|\sigma_{m}|11\rangle \|$ converges to $1/2$. Therefore, $\sigma_{m}$ converges to a private state by the recurrence protocol when \bea\| \langle 00|\sigma_{m}|11\rangle\| = \frac{\|\sigma_{1} - \sigma_{2}\|^{m}}{2\|\sigma_{1} + \sigma_{2}\|^{ m} + 2\|\sigma_{3} + \sigma_{4} \|^{ m}}\label{a0011}\eea can be arbitrarily close to $1/2$ as repetitions $m$ become very large. For such a convergence the shield states should fulfill that \bea \|\sigma_{1} +\sigma_{2}\| & = & \|\sigma_{1} -\sigma_{2}\|, \label{discond}\eea since it holds that $ \|\sigma_{1} +\sigma_{2} \| \geq \|\sigma_{1} - \sigma_{2} \| > \|\sigma_{3} + \sigma_{4} \|$, where the second inequality follows from the condition in (\ref{ent}). The condition (\ref{discond}) is in fact equivalent to the orthogonality, $\tr[\sigma_{1}\sigma_{2}] =0$ \cite{snu}. \\

\textbf{Proposition 1.} A secret key can be distilled from copies of the state $\rho_{ABA'B'}$ in (\ref{state}) using the recurrence protocol if shield sates
$\sigma_{1}$ and $\sigma_{2}$ in (\ref{state}) are orthogonal.\\

We now reconsider the bound entangled state in Ref. \cite{horodecki}
which takes the shield states as follows, \bea
\sigma_{1} & = & p ( \frac{\rho_{s}+\rho_{a}}{2} )^{\otimes l},
~~~~\sigma_{2} = p \rho_{s}^{\otimes l}, \nonumber \\
\sigma_{3,4} &=& (\frac{1}{2} - p)
(\frac{\rho_{s}+\rho_{a}}{2})^{\otimes l}, \label{taus} \eea where
$\rho_{a(s)}$ is the normalized $d$-dimensional projection
operator onto asymmetric(symmetric) space. The state
$\rho_{ABA'B'}$ remains positive under partial transpose if and only if $p \in (0,1/3]$
and $p\leq (1 + (d/(d-1))^{l})^{-1}$. From the condition in (\ref{discond}), it follows that a secret key can be distilled from copies of $\rho_{ABA'B'}$ only when $l$ becomes very large, since $\tr[\sigma_{1}\sigma_{2}] \propto 2^{-l}$.




\section{Distilling a secret key by the classical communication}

We now apply a classical key distillation protocol to distill a secret key from the states in (\ref{state}). The classical protocol involves in two-way communication, the advantage distillation plus one-way information reconciliation. It is worth to note here that, as we have mentioned in Sec. \ref{shiled}, when the key parts are measured in the computational basis, the ccq correlations are the same in both the original state $\rho_{ABA'B'}$ and the twisted one $U\rho_{ABA'B'}U^{\dagger}$ with any twisting operation $U$ \cite{pri}. Therefore, in order to analyze the key distillability of the original state, one can equivalently consider the ccq correlations in the twisted state.

To be explicit, suppose that a twisting operation is applied such that the key part results in the state in (\ref{p-state}). If the state in (\ref{p-state}) is separable, no secret key can be distilled since two honest parties share no secret correlations. The precondition for key distillation is that the state in (\ref{p-state}) is entangled. For the full generality, Eve is assumed to hold the purification of the state $\sigma_{AB}$ in (\ref{p-state}), \bea |\psi\rangle_{ABE} = \sum_{i} \sqrt{\lambda_{i}} |\phi_{i}\rangle_{AB} | E_{i}\rangle_{E}, \label{puri}\eea where $|E_{i}\rangle$ are orthogonal basis for Eve's Hilbert space. Alice and Bob measure their key parts in the computational basis and share the following probability distributions, \bea p_{AB}(i,j) = \tr[ |i,j\rangle_{AB}\langle i,j | \otimes I_{E}   |\psi\rangle_{ABE} \langle\psi|]. \eea For each measurement outcome $(i,j)$, Eve possesses the following state \bea |e_{i,j}\rangle_{E} = \frac{\langle i,j|_{AB}\otimes I_{E} |\psi\rangle_{ABE}}{\sqrt{p_{AB}(i,j)}}. \eea Then, the ccq correlations can be seen in the ccq state $\rho_{ABE}$ as follows, \bea \rho_{ABE} = \sum_{i,j} p_{AB}(i,j)|ij\rangle_{AB}\langle ij|\otimes |e_{ij}\rangle \langle e_{ij}|. \label{ccq}\eea

Sharing secret correlations, two honest parties now proceed to classical communication to distill a secret key. Here, we consider the advantage distillation that involves two-way classical communication \cite{Maurer CAD} plus one-way information reconciliation, which is known as the most tolerant classical key distillation protocol to date. The advantage distillation works as follows. Alice first generates a secret bit $s_{A}$ and computes a list $x_{i} = s_{A}+a_{i}$ with her measurement outcomes $a_{i}$, $i=1,\cdots,N$, and then announces $x_{i}$ through a public and authenticated channel so that Bob can also compute $b_{i}+x_{i}=y_{i}$. Bob will see that his resulting values are either all the same or not, and reply to Alice with \emph{acceptance} or \emph{rejection}, depending on the computation result. He says acceptance if all $y_{i}$ are the same, and proceeds to apply one-way error correction and privacy amplification to the accepted values. Otherwise, two honest parties leave from the failed values and perform the protocol with another values again. The advantage distillation can be understood as a post-selection processing to take bits having stronger correlations. Then, the standard one-way post-processing is applied. If a secret key is distilled after the one-way post-processing, then it is concluded that the original state $\rho_{ABA'B'}$ is key-distillable.

Key distillation from the states in (\ref{ccq}) using the advantage distillation followed by one-way post-processing has been completely analyzed in Refs. \cite{formula, BKR}. The key distillability condition is that, a secret key can be distilled from copies of the ccq state in (\ref{ccq}) if the ccq state satisfies that \bea |\langle e_{00} | e_{11}\rangle|^{2} > \frac{p_{AB}(0,1) + p_{AB}(1,0)}{p_{AB}(0,0) + p_{AB}(1,1)}. \eea In terms of the parameters $\lambda_{j}$ in (\ref{puri}), the condition is, $(\lambda_{1} - \lambda_{2})^{2} > (\lambda_{3}+\lambda_{4}) ( \lambda_{1} + \lambda_{2} ) $. Then, straightforwardly using the relation in (\ref{lam}), we arrive at the following proposition.\\

\textbf{Proposition 2.} A secret key can be distilled from copies of the state $\rho_{ABA'B'}$ using the advantage distillation followed by the one-way post-processing if shield sates satisfy the following condition, \bea \|\sigma_{1} - \sigma_{2} \|^{2} > \|\sigma_{3} +\sigma_{4} \| \|\sigma_{1} +\sigma_{2} \|. \label{form}\eea \\

The proposition $2$ also shows that, for a particular choice of separable shield states such that (\ref{discond}) holds, entanglement already implies that a secret key can be distilled. This is because, by the orthogonality condition (\ref{discond}), the entanglement
condition (\ref{ent}) coincides to the security condition in (\ref{form}). Note that, the security condition in (\ref{form}) is in general stronger than the entanglement condition in (\ref{ent}). Therefore, there exists a gap between entanglement condition and the key distillation condition.

We now compare two key distillability conditions in propositions 1 and 2. In fact, all quantum states which can be distilled to a secret key by the recurrence protocol can also be distilled to a secret key by the classical distillation protocol. Therefore, quantum states in much wider range are key-distillable.

\textbf{Proposition 3.} Let $S_{R}$ and $S_{C}$ denote the sets of
quantum states that can be distilled to secret by the recurrence protocol and by
the considered classical distillation, respectively. Then, it holds that $S_{R}
\subset S_{C}$. \\

\emph{Proof.} Suppose that $\rho_{ABA'B'}\in S_{R}$, meaning that the condition (\ref{discond}) is fulfilled. The classical distillation protocol is applied to those key-distillable
states. Since shield states fulfill (\ref{discond}), the security
condition (\ref{form}) coincides to the condition that
$\sigma_{AB}$ is entangled in (\ref{ent}). For key distillation, it is the precondition that $\sigma_{AB}$ is entangled, and therefore the security condition (\ref{form}) is fulfilled. This clearly shows that $S_{R}\subset S_{C}$. $\Box$\\

We now reconsider the example in Ref. \cite{hronew} with the classical distillation protocol. \\

\begin{figure}
  \includegraphics[width=8cm]{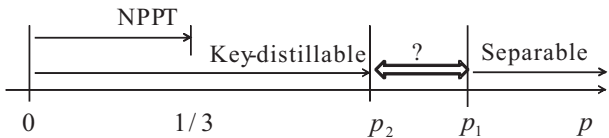}\\
  \caption{Those bounds of PPT and key-distillable in the example are shown. The range of separable means that the key part in (\ref{p-state}) is separable, meaning the range that no secret key can be distilled. }\label{gap}
\end{figure}

\emph{Example.} Let us revisit the state $\rho_{ABA'B'}$ with shield states in (\ref{taus}). Then, from its twisted state, the state $\sigma_{AB}$ is entangled if $ p\leq p_{1}$ where, for $j=1,2$, \bea p_{j}= \frac{1}{2}[(1-\frac{1}{2^{l}})^{j} +1]^{-1}. \label{pp} \eea The original state $\rho_{ABA'B'}$ remains positive under partial transpose if $p\leq 1/3$. From the key distillability condition in (\ref{form}), a secret key can be distileld from copeis of the state by the classical distillation protocol if $p > p_{2}$. This shows that the state $\rho_{ABA'B'}$ is key-distillable with shield states for all $l\geq 1$, contrast to the case that the recurrence protocol is applied, where a secret key is distilled only when $l$ is very large. Note that as $l$ goes very large, $p_{2}$ converges to $p_{1}$. All these are depicted in Fig. (\ref{gap}). \\

Based on the results from propositions 1 and 2, we also consider the example shown in Ref. \cite{hronew}, where a key-distillable bound entangled state in $4\otimes 4$ dimension is provided. Let us denote the state by $\rho_{ABA'B'}^{4\times 4}$. The state is of the form in (\ref{state}) with the following shield states \bea \sigma_{1} & = & \frac{q_{1}}{4} (|00\rangle\langle 00| + |\phi_{3}\rangle\langle\phi_{3}|), \nonumber \\  \sigma_{2} & = & \frac{q_{1}}{4} (|11\rangle\langle 11| + |\phi_{4}\rangle\langle\phi_{4}|) \nonumber \\
\sigma_{3} & = & \frac{q_{2}}{2} |\chi_{+}\rangle \langle \chi_{+}|,~~~\sigma_{4} =  \frac{q_{2}}{2} |\chi_{-}\rangle \langle \chi_{-}|\label{ex}\eea where \bea |\chi_{\pm} \rangle =  \frac{1}{2} (\sqrt{2 \pm \sqrt{2}} |00\rangle \pm \sqrt{2 \mp \sqrt{2}} |11\rangle). \nonumber \eea In Ref. \cite{hronew}, the ratio of $q_{1}$ and $q_{2}$ was found such that the state $\rho_{ABA'B'}^{4\times 4}$ provides ccq correlations having a positive secret key rate with the standard one-way protocol of error correction and privacy amplification.

The state $\rho_{ABA'B'}^{4\times 4}$ can actually be distilled to a secret key in a wider range of $q_{1}$ and $q_{2}$ if the classical distillation, advantage distillation plus one-way post-processing, is applied. First, note that the twisted state $\sigma_{AB}^{4\times 4}$ of state $\rho_{ABA'B'}^{4\times 4}$ (see, (\ref{p-state})) is entangled if $q_{1}> q_{2}$. Then, from propositions 1 and 2, the key-distillability can be seen explicitly as follows. Suppose that $q_{1}> q_{2}$. By the recurrence protocol, the state $\rho_{ABA'B'}^{4\times 4}$ can be approximated arbitrarily close to a private state since two states, $\sigma_{1}$ and $\sigma_{2}$, are orthogonal. Also, by the classical distillation and from the proposition 2, the state can be distilled to a secret key if the twisted state $\sigma_{AB}^{4\times 4}$ is entangled. Therefore, the state $\rho_{ABA'B'}^{4\times 4}$ is actually key-distillable in the much wider range, for $q_{1} > q_{2}$.

\section{When the white noise is added}

Finally, let us now consider the case that white noise is added to the original state $\rho_{ABA'B'}$, which is related with the case where a small error appears while or after identifying shared states by the LOCC scheme in \cite{pri}: \bea \rho_{ABA'B'}^{\epsilon} = (1-\epsilon)\rho_{ABA'B'} + \epsilon \openone_{ABA'B'}, \eea where $\openone_{ABA'B'}$ is the normalized identity operator. The state can be rewritten in a form (\ref{state}) with shield states $\sigma_{i}^{\epsilon}$ in the following \bea \sigma_{i}^{\epsilon} = (1-\epsilon)\sigma_{i} + \epsilon \openone_{A'B'}/4.\eea It is clear that shield states  $\sigma_{1}^{\epsilon}$ and $\sigma_{2}^{\epsilon}$ under the small perturbation do not satisfy the orthogonality even in the case that the original ones $\sigma_{1}$ and $\sigma_{2}$ might be orthogonal. Therefore, as the orthogonality is the key distillability condition for the case that the recurrence protocol is applied, a secret key cannot be distilled from states in (\ref{state}) by the recurrence protocol when the white noise is added to the states.

When the classical distillation protocol is applied, from the proposition $2$, a secret key can be distilled from copies of state, $\rho_{ABA'B'}^{\epsilon}$, if the following satisfies, \bea \| \sigma_{1} - \sigma_{2} \|^{2} > \| \sigma_{3} + \sigma_{4} \| \| \sigma_{1} + \sigma_{2} \| + \frac{\epsilon}{(1-\epsilon)^{2}}. \eea Therefore, the classical distillation protocol can still work although the white noise is added to the original state. As an example, we apply the above security condition to the example in Ref. \cite{horodecki}: a secret key can be distilled if \bea p> \frac{p_{2}}{2}  [1+ (1-\frac{2}{p_{2}} \frac{\epsilon}{(1-\epsilon)^{2}})^{1/2}], \label{ncon} \eea  where $p_{2}$ is given in (\ref{pp}).

\section{Conclusion}

To summarize, we have investigated the key distillability of shielded two-qubit states using the quantum protocol, the recurrence protocol, and the classical distillation protocol, the advantage distillation plus one-way post-processing. We then derive the security conditions, respectively, and show that entangled states in a much wider range are in fact key-distillable. We also reconsider known key-distillable bound entangled states shown in Refs. \cite{horodecki} and \cite{hronew}. It is also shown that, for a particular choice of shield states such that $\sigma_{1}$ and $\sigma_{2}$ in (\ref{state}) are orthogonal, entanglement immediately implies that a secret key can be distilled. Fianlly, the classical distillation is shown to be robust when the white noise is added to original quantum state.

This work is supported by the Korea Research Foundation Grant funded by the Korean Government, KRF-2008-313-C00185, and the IT R$\&$D program of MKE/IITA (2008-F-035-01).


\begin{thebibliography}{99}
\bibitem{skey}
A secret key has the following probability distribution, i)perfect
correlation between two honest parties, $ p_{AB}(0,0) =
p_{AB}(1,1) = 1/2$ and ii) independence against an eavesdropper, $
p_{ABE} = p_{AB} p_{E}$.
\end{thebibliography}
\end{document}